\documentclass[12pt]{iopart}
\usepackage{iopams,cite,graphicx,setstack}

\newcommand{\dd}{{\rm d}}
\newcommand{\ee}{{\rm e}}

\begin{document}

\title[Anisotropic collective flow of a Lorentz gas]%
  {Anisotropic collective flow of a Lorentz gas}

\author{\underline{N Borghini} and C Gombeaud}

\address{Fakult\"at f\"ur Physik, Universit\"at Bielefeld, 
  Postfach 100131, D-33501 Bielefeld, Germany}

\begin{abstract}
Analytical results for the anisotropic collective flow of a Lorentz gas of 
massless particles scattering on fixed centres are presented. 
\end{abstract}



A remarkable feature of nucleus-nucleus collisions is the anisotropy of the 
particle emission pattern in the plane transverse to the collision axis: the
transverse momentum distribution of outgoing particles reads
\begin{equation}
\label{def-vn}
\frac{\dd^2N}{\dd^2\bf p} = \frac{1}{2\pi}\frac{\dd N}{p_T\,\dd p_T} \Bigg[
  1 + \sum_{n=1}^\infty 2v_n(p_T)\cos n(\varphi-\Phi_n) \Bigg],
\end{equation}
with $v_n(p_T)$ the ``anisotropic flow'' coefficients, $\varphi$ the azimuth of 
transverse momentum ${\bf p}$ and $\Phi_n$ the event-by-event varying reference 
angle for the $n$th flow harmonic.
Hereafter we shall neglect fluctuations, and all $\Phi_n$ will coincide with 
the $x$-axis.

The experiment-driven focus of theoretical studies in the recent years has been 
on anisotropic flow for matter close to equilibrium. 
Here, we want to investigate the opposite case when particles undergo very few
rescatterings, so that their evolution can meaningfully be described by a 
kinetic equation of the Boltzmann type. 
We specifically aim at obtaining analytical results---similar to those derived 
in \cite{Heiselberg:1998es}---which allow us to clearly identify qualitative 
behaviours together with their possible origins.

As a further simplification, we consider the anisotropic flow of a ``Lorentz 
gas'' of massless particles diffusing on infinitely massive particles. 
This constitutes a regular yet much simpler limiting case for the scattering of 
light particles on massive ones~\cite{Borghini:2010hy}.

We wish to stress that the {\em qualitative\/} features which we derive in the 
following Sections are to our eyes more robust and thereby more important than 
the quantitative results.
The model of a Lorentz gas may have little relevance for the phenomenology of 
heavy-ion collisions, yet it allows us to exemplify how in a more realistic 
description one should naturally expect
\begin{itemize}
\item the mixing of different flow harmonics;
\item the evolution of anisotropic flow in the absence of spatial asymmetry 
  when some flow is already present;
\item the non-monotonic time evolution of anisotropic flow. 
\end{itemize}
Our simple model also shows that such complex qualitative behaviours are not 
the exclusive privilege of approaches assuming many rescatterings like 
(dissipative) fluid dynamics, but can appear quite generally, and play a role 
either at very early times~\cite{Krasnitz:2002ng,Broniowski:2008qk} or around
the kinetic freeze-out, as well as for the anisotropic flow of fragile states.

\section{Expansion of a Lorentz gas}
\label{s:intro}

Consider a gas of $N$ massless particles, described by the distribution density 
$f(t, {\bf r}, {\bf p})$, that scatter elastically with the differential cross 
section $\sigma_\dd$ on a distribution $n_c({\bf r})$ of fixed scattering 
centres. 
We shall assume that the problem is two-dimensional, i.e. we focus on the 
transverse dynamics of the gas, so that $\sigma_\dd$ has the dimension of a 
length.

$f$ then obeys the Boltzmann--Lorentz kinetic equation
\begin{equation}
\label{Lorentz}
\fl \partial_t f(t, {\bf r}, {\bf p}) + 
  {\bf v}\cdot{\bf\nabla}_{\bf r}f(t, {\bf r}, {\bf p}) = 
  n_c({\bf r})\,c \!\int\!\dd\Theta\,\sigma_\dd(\Theta)\,
    \big[ f(t, {\bf r}, {\bf p}') - f(t, {\bf r}, {\bf p}) \big],
\end{equation}
with ${\bf v}$ the particle velocity and $\Theta$ the scattering angle of the 
diffusing particle. 

Integrating \Eref{Lorentz} over space, the gradient term disappears, and one  
finds the evolution equation for the particle momentum distribution 
$\dd^2N/\dd^2{\bf p}$.
The latter can then be multiplied by $\cos n\varphi$, with $\varphi$ the azimuth
of ${\bf p}$, and averaged over $\varphi$, yielding the evolution equation for
the anisotropic flow harmonic $v_n$.

At vanishing cross section, the solutions to \Eref{Lorentz} are the 
{\it free-streaming solutions\/}
\begin{equation}
\label{f_free}
f^{(0)}(t, {\bf r}, {\bf p}) = f^{(0)}(0, {\bf r}-{\bf v}t, {\bf p}),
\end{equation}
which are entirely determined by the initial distribution at $t=0$.

In the following, we study small deviations $f=f^{(0)}+f^{(1)}$ with 
$|f^{(1)}|\ll f^{(0)}$ to these solutions---which corresponds to considering 
very few scatterings per particle---by injecting the free-streaming solution in 
the collision integral in \Eref{Lorentz}. 
Introducing the total elastic cross section
\[
\sigma_{\rm el.} \equiv \int\!\dd\Theta\,\sigma_\dd(\Theta)
\]
and the {\em unintegrated kernel\/}
\begin{equation}
\label{C(X,p)}
{\cal C}({\bf X},{\bf p}) \equiv \int\!\dd^2{\bf r}\,n_c({\bf r})\,
  f^{(0)}(0,{\bf r}-{\bf X},{\bf p}),
\end{equation}
one finds
\begin{equation}
\label{d2N/dpdt_bis}
\partial_t\Bigg[\frac{\dd^2N}{\dd^2{\bf p}}(t,{\bf p})\Bigg] \simeq 
  c\Bigg[\int\!\dd\Theta\,\sigma_\dd(\Theta)\,{\cal C}({\bf v}'t,{\bf p}') - 
  \sigma_{\rm el.}\,{\cal C}({\bf v}t,{\bf p}) \Bigg].
\end{equation}
The scattering rate at time $t$ is given by
\begin{equation}
\label{Gamma(t)}
\Gamma(t) = \int\!\dd^2{\bf p}\,\dd^2{\bf r}\,
  n_c({\bf r})\,f(t,{\bf r},{\bf p})\,\sigma_{\rm el.} c \approx
  \sigma_{\rm el.} c\!\int\!\dd^2{\bf p}\ {\cal C}({\bf v}t,{\bf p}).
\end{equation}
Integrated over time, this rate gives the total number of rescatterings 
${\cal N}_{\rm scat.}$, which for the consistency of our approach should be 
small.

For the density of scattering centres and the density distribution of diffusing
particles  at the initial time $t=0$, we assume Gaussian profiles in position 
space
\begin{equation}
\label{nc(r)}
n_c({\bf r}) = \frac{N_c}{2\pi R_xR_y}
  \exp\!\left(\!-\frac{x^2}{2R_x^2}-\frac{y^2}{2R_y^2} \right),
\end{equation}
with $N_c$ the total number of scattering centres, and
\begin{equation}
\label{f(0,r,p)}
f^{(0)}(0,{\bf r}, {\bf p}) = \frac{N\tilde f({\bf p})}{4\pi^2 R_xR_y}\,
  \exp\!\left(\!-\frac{x^2}{2R_x^2}-\frac{y^2}{2R_y^2} \right),
\end{equation}
where the initial momentum distribution $\tilde f({\bf p})$ is normalized to 
$2\pi$, so that the integral of $f(0,{\bf r}, {\bf p})$ over space and momentum 
yields the total number of diffusing particles. 
For the sake of simplicity we consider identical radii $R_x$, $R_y$ for both
distributions. 
Let
\[
R_x^2\equiv\frac{R^2}{1+\epsilon},\quad R_y^2\equiv\frac{R^2}{1-\epsilon}.
\]

With the initial profiles~\eref{nc(r)} and \eref{f(0,r,p)}, the unintegrated 
kernel~\eref{C(X,p)} reads
\begin{equation}
\label{C(X,p)_2D}
{\cal C}({\bf X},{\bf p}) = 
\frac{N_c N\tilde f({\bf p})}{8\pi^2 R^2}\sqrt{1-\epsilon^2}\,\exp\!
  \left[\!-\frac{X^2(1+\epsilon)+Y^2(1-\epsilon)}{4R^2}\right],
\end{equation}
with ${\bf X}=(X,Y)$.

\section{Isotropic initial momentum distribution, isotropic cross section}
\label{s:2D_all-isotropic}

Let us first consider the simplest case of an isotropic initial momentum 
distribution $\tilde f({\bf p})\equiv \tilde f_0(p_T)$ as well as an 
isotropic differential cross section $\sigma_\dd$.
The latter can then be replaced by $\sigma_{\rm el.}/2\pi$ and taken out of the 
gain term in \Eref{d2N/dpdt_bis}.
Note that our normalization choice for $\tilde f({\bf p})$ is equivalent to 
\[
\int_0^\infty\!\tilde f_0(p_T)\,p_T\,\dd p_T = 1.
\]

Let $\varphi$ (resp.\ $\varphi'$) denote the azimuth of ${\bf p}$ (resp.\ 
${\bf p}'$) with respect to the direction of the $x$-axis of the scattering 
centre distribution $n_c$. 
Then
\begin{equation}
\label{C(vt,p)_2D}
{\cal C}({\bf v}t,{\bf p}) = \frac{N_c N\tilde f_0(p_T)}{8\pi^2 R^2}\,
  \sqrt{1-\epsilon^2}\,\ee^{-c^2 t^2/4R^2}
   \exp\!\left(\!-\frac{c^2 t^2}{4R^2}\epsilon\cos 2\varphi\right),
\end{equation}
and an analogous equation for ${\cal C}({\bf v}'t,{\bf p}')$.

This expression is readily integrated over ${\bf p}$, yielding the rate
\begin{equation}
\label{Gamma(t)_2D}
\Gamma(t) = \frac{N_c N\sigma_{\rm el.} c}{4\pi R^2}\,\sqrt{1-\epsilon^2}\,
  \ee^{-c^2 t^2/4R^2}\,I_0\Bigg(\frac{c^2 t^2}{4R^2}\epsilon\Bigg), 
\end{equation}
with $I_0$ the modified Bessel function of the first kind.
Integrating from $t=0$ to infinity gives the total number of rescatterings 
over the evolution:
\begin{equation}
\label{Ncoll_2D}
{\cal N}_{\rm scat.} = \frac{N_c N\sigma_{\rm el.}}{2\pi^{3/2}R}\,
  \sqrt{1-\epsilon}\, K\Bigg(\frac{2\epsilon}{1+\epsilon}\Bigg),
\end{equation}
where $K$ denotes the complete elliptic integral of the first kind.
At given $N_c$, $N$, $R$ and $\sigma_{\rm el.}$, this number of rescatterings is
maximal for $\epsilon=0$: 
one can thus fix the average number of rescatterings per diffusing particle 
${\cal N}_{\rm scat.}/N$ at some small value in central collisions---which 
amounts to fixing the ratio $N_c\sigma_{\rm el.}/R$---and ensure a small number 
of rescatterings over all centralities.

The time evolution of the $n$th anisotropic flow harmonic follows from
\begin{equation}
\label{dvn/dt_2D}
\fl\partial_t v_n(t,p) \equiv \partial_t\left[ \frac{%
  \int_0^{2\pi}\!\dd\varphi\,\frac{\dd^2N}{\dd^2{\bf p}}(t,{\bf p})\,
    \cos n\varphi}{%
  \int_0^{2\pi}\!\dd\varphi\,\frac{\dd^2N}{\dd^2{\bf p}}(t,{\bf p})} \right] =
  \frac{1}{N\tilde f_0(p_T)} \int_0^{2\pi}\!\dd\varphi\,
    \partial_t\Bigg[\frac{\dd^2N}{\dd^2{\bf p}}(t,{\bf p})\Bigg]\cos n\varphi,
\end{equation}
where in writing the denominator we have used the fact that elastic collisions 
on fixed scattering centres leave the momentum modulus $p_T$ unchanged, so that 
the momentum spectrum is actually independent of time. 
The integrand in the rightmost expression can be rewritten using 
\Eref{d2N/dpdt_bis} and corresponds to the collision integral. 
The unintegrated kernel involved is given by \Eref{C(vt,p)_2D}. 
First, the gain term depends on $\varphi'$, not on $\varphi$, and thus does not
contribute to $\partial_t v_n(t,p_T)$. 
Then, the loss term yields
\begin{equation}
\label{dvn/dt_2D_loss}
\fl \partial_t v_n(t,p_T)\Big|_{\rm loss} = 
\left\{\!
\begin{array}{ll}
  0 & \mbox{for odd }n;\\[2mm]
  \displaystyle (-1)^{1+n/2} \frac{N_c\sigma_{\rm el.}c}{4\pi R^2}\,
  \sqrt{1-\epsilon^2}\,\ee^{-c^2 t^2/4R^2}\,
    I_{\!\frac{n}{2}}\!\Bigg(\frac{c^2t^2}{4R^2}\epsilon\Bigg) & 
    \mbox{for even }n.
\end{array}\right. 
\end{equation}
Note that a factor of 2 is missing in the denominator of the equation as written
in \cite{Borghini:2010hy}.

All $v_{4n+2}$ coefficients, in particular $v_2$, are increasing with time,
while the Fourier harmonics $v_{4n}$ are decreasing. 
Since the coefficients vanish at $t=0$---the momentum distribution $\tilde f$ 
is isotropic---one deduces for instance $v_2>0$, but $v_4<0$: 
this reflects the alternating signs of the corresponding moments (in position 
space) of the initial Gaussian profiles. 

Integrating \Eref{dvn/dt_2D_loss} over time from 0 to $t$ yields the time 
dependence of the Fourier coefficients $v_n(t,p_T)$. 
At early times $|v_n(t,p_T)|\propto t^{n+1}$, while at late times one finds for 
even $n$~\cite[formula 2.15.3(2)]{Prudnikov2}
\begin{eqnarray}
\label{vn(p)_2D}
\fl v_n(p_T) &\equiv\lim_{t\to\infty} v_n(t,p_T) \cr
\fl &= (-1)^{n/2+1}\frac{N_c\sigma_{\rm el.}\sqrt{1-\epsilon^2}}{4\sqrt{\pi}R}\,
  \frac{(n-1)!!}{2^n\big(\frac{n}{2}\big)!}\,
  _2F_1\bigg(\frac{n+1}{4},\frac{n+3}{4};\frac{n}{2}+1;\epsilon^2\bigg)\,
  \epsilon^{n/2},
\end{eqnarray}
where $(2k-1)!! = 1 \cdot 3 \cdots (2k-1)$ if $k\geq 1$, 1 if $k = 0$, while 
$_2F_1$ denotes the Gaussian hypergeometrical function.
For $n=2$ (resp.\ $n=4$), this formula reduces to Equation (C4) (resp.\ (C5)) 
of~\cite{Borghini:2010hy}.
Interestingly, $v_n(p_T)$ scales as $\epsilon^{n/2}$ for small eccentricities.

\begin{figure}
  \centerline{\includegraphics*[width=0.7\linewidth]{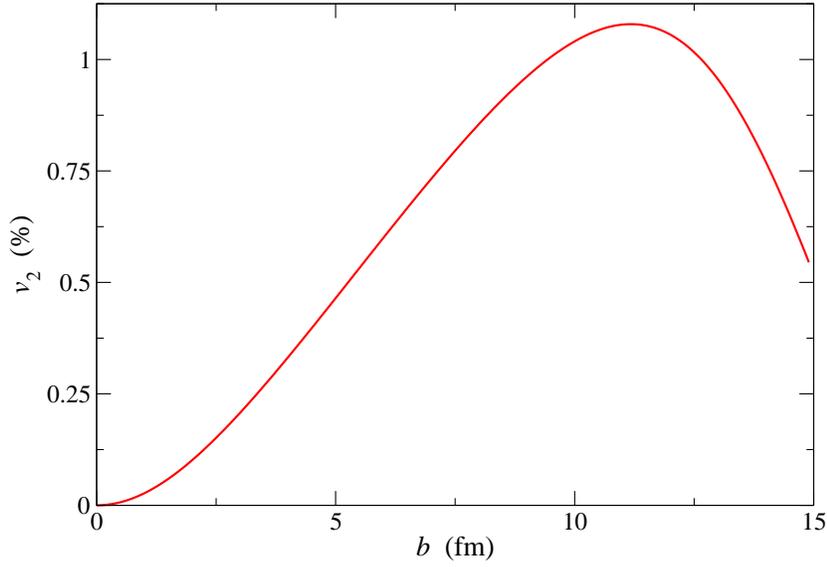}}
  \caption{\label{fig:v2(b)}Centrality dependence of $v_2$ (here, for an 
    evolution with on average 0.1 collision per particle).}
\end{figure}

In Figure~\ref{fig:v2(b)}, we show the dependence of $v_2$ on impact parameter 
$b$ in Pb-Pb collisions, where the eccentricity $\epsilon$ is related to $b$ 
through the Glauber optical model, assuming that the eccentricity dependence of
$v_2$ is given by \Eref{vn(p)_2D} with $n=2$.

\section{Anisotropic initial momentum distribution, isotropic cross section}
\label{s:2D_f(p)_anisotropic}

We now allow for the possibility that the expanding gas possess an initially 
anisotropic momentum distribution, which we describe by introducing its Fourier 
series
\begin{equation}
\label{f(p)_anisotropic}
\tilde f({\bf p}) = \tilde f_0(p_T)\Bigg[ 1 + 2\sum_{k=1}^\infty
  \big( w_{k,c}\cos k\varphi + w_{k,s}\sin k\varphi \big) \Bigg].
\end{equation}
The Fourier coefficients $w_{k,c}$, $w_{k,s}$ could generally depend on $p_T$, 
yet we shall hereafter leave this dependence aside.
The coefficients $w_{n,c}$ of the cosine harmonics correspond to the ``usual'' 
anisotropic flow coefficients, taken at the initial time, $v_n(t=0,p_T)$.
As in the previous Section, the differential cross section is taken to be 
isotropic. 
\smallskip

The initial momentum distribution~\eref{f(p)_anisotropic} gives for the 
unintegrated kernel 
\begin{eqnarray}
\label{C(vt,p)_2D_f(p)_anisotropic}
\fl {\cal C}({\bf v}t,{\bf p}) = \frac{N_c N\tilde f_0(p_T)}{8\pi^2 R^2} &
  \sqrt{1-\epsilon^2}\ \ee^{-c^2 t^2/4R^2} \cr
\fl &\times \exp\!\left(\!-\frac{c^2 t^2}{4R^2}\epsilon\cos 2\varphi\right)\! 
  \Bigg[ 1 + 2\sum_{k=1}^\infty \big( w_{k,c}\cos k\varphi + 
     w_{k,s}\sin k\varphi \big) \Bigg].
\end{eqnarray}
With this kernel, the scattering rate is given by
\[
\fl\Gamma(t) = \frac{N_c N\sigma_{\rm el.} c}{4\pi R^2}\sqrt{1-\epsilon^2}\ 
  \ee^{-c^2 t^2/4R^2}\,\Bigg[ I_0\Bigg(\frac{c^2 t^2}{4R^2}\epsilon\Bigg) +
    2\sum_{q\geq 1} (-1)^q w_{2q,c}
    I_q\Bigg(\frac{c^2 t^2}{4R^2}\epsilon\Bigg) \Bigg],
\]
and the total number of rescatterings, which has to be kept small, by
\begin{eqnarray*}
\label{Ncoll_2D_f(p)_anisotropic}
\fl {\cal N}_{\rm scat.} = \frac{N_c N\sigma_{\rm el.}}{4\sqrt{\pi}R}\,
  \sqrt{1-\epsilon^2}\,\Bigg[\,\! & 
    _2F_1\Bigg(\frac{1}{4},\frac{3}{4};1;\epsilon^2\Bigg) \cr
\fl & + 
   \sum_{q\geq 1} (-1)^q w_{2q,c}\frac{(2q\!-\!1)!!}{2^{2q-1}q!}\,
     _2F_1\Bigg(\frac{2q\!+\!1}{4},\frac{2q\!+\!3}{4};q\!+\!1;\epsilon^2\Bigg)\,
     \epsilon^q \Bigg].
\end{eqnarray*}

The unintegrated kernel~\eref{C(vt,p)_2D_f(p)_anisotropic} also allows one to 
compute the time derivative of the anisotropic flow coefficient $v_n$. 
As in Section~\ref{s:2D_all-isotropic}, the gain term of the collision integral 
does not contribute to $\partial_t v_n(t,p_T)$, whereas the contribution of the 
loss term follows from multiplying \Eref{C(vt,p)_2D_f(p)_anisotropic} with 
$\cos n\varphi$ and then integrating over $\varphi$.
The only terms from the sum over $k$ that result in a non-vanishing integral are
those in $\cos k\varphi$ with $k$ of the same parity as $n$, so that $n-k$ and 
$n+k$ are even.
The isotropic part of the momentum distribution only contributes when $n$ is 
even, as in \Eref{dvn/dt_2D_loss}.

\noindent All in all, one finds
\numparts
\begin{eqnarray}
\fl\partial_t& v_{2m}(t,p_T)\Big|_{\rm loss} \!= 
  (-1)^{m+1}\frac{N_c\sigma_{\rm el.} c}{4\pi R^2}&\sqrt{1-\epsilon^2}\ 
  \ee^{-c^2 t^2/4R^2} \Bigg\{ 
    I_m\Bigg(\frac{c^2 t^2}{4R^2}\epsilon\Bigg) \cr
\fl & &  \!\!\!+\! 
    \sum_{q\geq 1}(-1)^q w_{2q,c} \Bigg[
      I_{m+q}\Bigg(\frac{c^2 t^2}{4R^2}\epsilon\Bigg)\! +
      I_{m-q}\Bigg(\frac{c^2 t^2}{4R^2}\epsilon\Bigg) \! \Bigg] \!\Bigg\}, 
\label{dvn/dt_2D_f(p)_anisotropic_loss_even}
\end{eqnarray}
\begin{eqnarray}
\fl\partial_t& v_{2m+1}(t,p_T)\Big|_{\rm loss} \!= 
  (-1)^{m+1}&\frac{N_c\sigma_{\rm el.} c}{4\pi R^2}\sqrt{1-\epsilon^2}\ 
  \ee^{-c^2 t^2/4R^2} \cr
\fl & & \times \sum_{q\geq 1}(-1)^q w_{2q-1,c} 
  \Bigg[ I_{m+q}\Bigg(\frac{c^2 t^2}{4R^2}\epsilon\Bigg)\! +
    I_{m-q}\Bigg(\frac{c^2 t^2}{4R^2}\epsilon\Bigg) \Bigg].
\label{dvn/dt_2D_f(p)_anisotropic_loss_odd}
\end{eqnarray}
\endnumparts

Let us shortly discuss these results, focusing first on the short-time 
behaviour.
Taking $m=1$, \Eref{dvn/dt_2D_f(p)_anisotropic_loss_even} gives for the 
evolution of elliptic flow
\[
\fl \partial_t v_2(t,p_T) \sim \frac{N_c\sigma_{\rm el.} c}{4\pi R^2}
  \sqrt{1-\epsilon^2}\left[-w_{2,c} + 
    \frac{c^2}{8R^2}\big(\epsilon+2w_{2,c}+\epsilon w_{4,c}\big) t^2 + \Or(t^4) 
  \right]\ \mbox{ for }t\ll\frac{R}{c}.
\]
That is, in the presence of a positive initial elliptic flow, $v_2(t,p_T)$ 
first decreases (linearly), before it starts increasing:
 since more particles are emitted in-plane than out-of-plane, there are more 
particles ``lost'' at $\varphi=0$ or 180$^{\rm o}$ than at $\pm 90^{\rm o}$. 
On the contrary, a negative initial $w_{2,c}=v_2(0,p_T)$ accelerates the initial
increase of elliptic flow, while the later behaviour depends on the sign of 
$\epsilon(1+w_{4,c})+2w_{2,c}$. 
In either case, $v_2$ evolves even for vanishing eccentricity $\epsilon=0$, 
which is quite a nontrivial finding. 
One also sees that $v_2$ is influenced by the presence of any finite initial 
$w_{4,c}=v_4(0,p_T)$.

For odd harmonics $n$, \Eref{dvn/dt_2D_f(p)_anisotropic_loss_odd} shows that 
some finite $v_n(t,p_T)$ can develop if and only if there exist at least one 
non-vanishing odd harmonic at $t=0$. 
In the case of directed flow $v_1$ for instance, \Eref{%
  dvn/dt_2D_f(p)_anisotropic_loss_odd} with $m=0$ gives
\begin{eqnarray*}
\partial_t v_1(t,p_T) &= 
  -\frac{N_c\sigma_{\rm el.} c}{2\pi R^2}\sqrt{1-\epsilon^2}\,
  \ee^{-c^2 t^2/4R^2} \sum_{q\geq 1}(-1)^q w_{2q-1,c} 
    I_q\Bigg(\frac{c^2 t^2}{4R^2}\epsilon\Bigg) \\
 &\sim \frac{N_c\sigma_{\rm el.} c^3}{16\pi R^4}\sqrt{1-\epsilon^2}\,
  w_{1,c}\epsilon t^2 + \Or(t^4)\ \mbox{ for }t\ll\frac{R}{c},
\end{eqnarray*}
whereas for triangular flow $v_3$, one finds
\[
\fl \partial_t v_3(t,p_T) \sim \frac{N_c\sigma_{\rm el.} c}{4\pi R^2}
  \sqrt{1-\epsilon^2} \left[-w_{1,c} + 
    \frac{c^2}{8R^2}\big(w_{3,c}\epsilon+2w_{1,c}\big) t^2  + \Or(t^4)\right]\ 
  \mbox{ for }t\ll\frac{R}{c}.
\]
Thus, $v_3$ evolves even in the absence of any ``triangularity'' in the 
collision geometry---in obvious similarity to the evolution of $v_2$ for 
$\epsilon=0$. 
Additionally, $\partial_t v_3(t,p_T)$ again illustrates the mixing of different 
harmonics present in 
Equations~\eref{dvn/dt_2D_f(p)_anisotropic_loss_even}--\eref{%
  dvn/dt_2D_f(p)_anisotropic_loss_odd}.

The latter can be integrated from $t=0$ to $\infty$.
One in particular gets
\begin{eqnarray*}
\fl v_2(p_T) = w_{2,c} + &
  \frac{N_c\sigma_{\rm el.}\sqrt{1\!-\!\epsilon^2}}{16\sqrt{\pi}R}\,\Bigg\{ 
    {}_2F_1\Bigg(\frac{3}{4},\frac{5}{4};2;\epsilon^2\Bigg)\epsilon \cr
\fl & \hspace{1cm}+ \sum_{q\geq 1}\!\Bigg(\frac{-1}{4}\Bigg)^{\!\!q} w_{2q,c}
   \Bigg[ \frac{(2q\!+\!1)!!}{(q\!+\!1)!}\, 
   _2F_1\Bigg(\frac{2q\!+\!3}{4},\frac{2q\!+\!5}{4};q\!+\!2;\epsilon^2\Bigg)
   \epsilon^{q+1} \cr
\fl & \hspace{4.5cm}+ 
   16\frac{|2q\!-\!3|!!}{(q\!-\!1)!}\, 
   _2F_1\Bigg(\frac{2q\!-\!1}{4},\frac{2q\!+\!1}{4};q;\epsilon^2\Bigg)
   \epsilon^{q-1} 
 \Bigg] \!\Bigg\}.
\end{eqnarray*}
The initial elliptic flow $w_{2,c}=v_2(t\!=\!0,p_T)$ breaks the linear scaling 
of $v_2$ with eccentricity at small $\epsilon$ both trivially as well as 
through its influence on the anisotropic flow developed in the rescatterings: 
\[
v_2(p_T) \underset{\epsilon\ll 1}{\sim} 
  \Bigg( 1 - \frac{N_c\sigma_{\rm el.}}{4\sqrt{\pi}R} \Bigg)v_2(t\!=\!0,p_T) + 
  \frac{N_c\sigma_{\rm el.}}{16\sqrt{\pi}R}\big[ 1 + 3v_4(t\!=\!0,p_T) \big]\,
    \epsilon.
\]
Again, we find the mixing of different harmonics as well as an evolving $v_2$ 
at $\epsilon=0$. 
Note that the ratio $N_c\sigma_{\rm el.}/4\sqrt{\pi}R$ necessarily takes a 
small value when the mean number of rescatterings per particle is small. 
Accordingly, $v_2(p_T)$ does not differ much from its initial value, which is 
normal within our few-rescatterings approach.

\section{Isotropic initial momentum distribution, anisotropic cross section}
\label{s:2D_sigma(Theta)}

We now come back to an isotropic initial momentum distribution, but consider
the case of an anisotropic differential cross section $\sigma_\dd(\Theta)$. 
The latter can generally be expanded as a Fourier series. 
The first harmonic in the expansion describes an asymmetry between forward and 
backward scattering, the former being favored if the corresponding coefficient 
is positive. 
Then, the second harmonic accounts for increased or suppressed scattering at 
$\pm 90^{\rm o}$ with respect to 0 or 180$^{\rm o}$.
Higher harmonics describe less natural behaviours, which we shall not consider 
in the following. 
Additionally, we assume that the interaction preserves parity, so that sine
harmonics vanish. 
We thus restrict ourselves to a differential cross section given by
\begin{equation}
\label{sigma(Theta)}
\sigma_\dd(\Theta) = \frac{\sigma_{\rm el.}}{2\pi}
  \big( 1 + 2\varsigma_1\cos\Theta + 2\varsigma_2\cos2\Theta \big).
\end{equation}
Note that the coefficients $\varsigma_1$ and $\varsigma_2$ are not totally
arbitrary, since $\sigma_\dd$ must remain non-negative when $\Theta$ spans the 
range $[0,2\pi]$: one for example easily checks that, irrespective of the value 
of $\varsigma_1$, one should have $|\varsigma_2|\leq \frac{1}{2}$.

The anisotropy of the differential cross section does not affect the scattering 
rate nor the resulting total number of rescatterings, which are thus given by 
Equations~\eref{Gamma(t)_2D} and \eref{Ncoll_2D}. 
The loss term of the collision integral relies on the total elastic cross 
section and is thus the same as in Section~\ref{s:2D_all-isotropic}: it still 
yields a contribution to $\partial_t v_n(t,p_T)$ given by the right-hand side of
\Eref{dvn/dt_2D_loss}.

On the other hand, the gain term of the collision integral now gives a 
non-vanishing contribution, since $\varphi'$ is no longer arbitrary, but related
to $\varphi$ through $\varphi'=\varphi+\Theta$, with a non-uniform distribution 
in $\Theta$.
Inspecting Equations~\eref{d2N/dpdt_bis}, \eref{C(vt,p)_2D} and \eref{dvn/dt_2D}
together with the differential cross section~\eref{sigma(Theta)}, the 
contribution to $\partial_t v_n(t,p)$ of the gain term reads
\begin{eqnarray*}
\fl\partial_t v_n(t,p_T)\Big|_{\rm gain} \!= &
  \frac{N_c\sigma_{\rm el.} c}{8\pi^2R^2}\,\sqrt{1-\epsilon^2}\ 
  \ee^{-c^2 t^2/4R^2} \ & \\
\fl & \times\!\int_0^{2\pi}\!\dd\varphi\,\cos n\varphi\,\Bigg\{\!
   \int_0^{2\pi}\!\frac{\dd\varphi'}{2\pi}\, & 
   \!\exp\!\left(\!-\frac{c^2 t^2}{4R^2}\epsilon\cos 2\varphi'\right) \\
\fl & & \!\times\!\Big[ 1 + 2\varsigma_1\cos(\varphi'\!-\!\varphi) + 
      2\varsigma_2\cos2(\varphi'\!-\!\varphi) \Big] \!\Bigg\}.
\end{eqnarray*}
Irrespective of the value of $n$, the $\varsigma_1$ term leads to a vanishing 
integral over $\varphi'$, while the integrals of the constant and $\varsigma_2$ 
terms yield modified Bessel functions, so that the expression between curly 
brackets equals
\[
I_0\Bigg(\frac{c^2 t^2}{4R^2}\epsilon\Bigg) - 
  2\varsigma_2 I_1\Bigg(\frac{c^2 t^2}{4R^2}\epsilon\Bigg)\cos 2\varphi.
\]

In turn, the remaining integral over $\varphi$ is trivial and yields for $n=2$
\[
\partial_t v_2(t,p_T)\Big|_{\rm gain} = -\frac{N_c\sigma_{\rm el.} c}{4\pi R^2}
  \sqrt{1-\epsilon^2}\,\ee^{-c^2 t^2/4R^2}\,
  I_1\Bigg(\frac{c^2 t^2}{4R^2}\epsilon\Bigg) \varsigma_2,
\]
while it vanishes for $n\neq 2$, i.e.\ the gain term only contributes to the 
second harmonic of anisotropic flow, that is elliptic flow.
Putting the gain and loss terms together, one eventually obtains after 
integrating over time
\begin{equation}
\label{v2(p)_2D_sigma(Theta)}
v_2(p_T) = (1-\varsigma_2)
  \frac{N_c\sigma_{\rm el.}\sqrt{1-\epsilon^2}}{16\sqrt{\pi}R}\,
  _2F_1\Bigg(\frac{3}{4},\frac{5}{4};2;\epsilon^2\Bigg)\,\epsilon,
\end{equation}
while $v_n(p_T)$ for even $n\neq 2$ remains given by \Eref{vn(p)_2D}.
Thus, an increased (resp.\ decreased) scattering probability at 
$\pm 90^{\rm o}$, as found e.g.\ in collisions of identical bosons (resp.\ 
fermions)---which is obviously not the case of the colliding particles in our 
Lorentz-gas model---, leads to a larger (resp.\ smaller) $v_2$.
\medskip

Eventually, one can mix the various ingredients together and consider an 
anisotropic differential cross section together with some initial anisotropic 
flow. 
In that case, the $\varsigma_1$ coefficient starts playing a role when combined 
with non-vanishing initial $w_{n,c}$, while $\varsigma_2$ will affect further 
flow harmonics besides $v_2$.

\ack
C G acknowledges support from the Deutsche Forschungsgemeinschaft under grant 
GRK 881.

\end{document}